%% file: main.tex
\documentclass[10pt, journal, compsoc]{IEEEtran}
\usepackage{cite}
\usepackage{amsmath,amssymb,amsfonts}
\usepackage{algorithmic}
\usepackage{graphicx}
\usepackage{textcomp}
\usepackage{todonotes}
\usepackage{url}
\usepackage[hidelinks]{hyperref}
\usepackage{enumitem}
\usepackage{xcolor}
\usepackage{subcaption}
\usepackage{multirow}
\usepackage{array}
\usepackage{tabulary}
\usepackage{url}

\newcommand{\System}{\textsc{Priveshield}}

\begin{document}
\sloppy

\title{PriveShield: Enhancing User Privacy Using Automatic Isolated Profiles in Browsers}
\author{Seyed Ali Akhavani*,
Engin Kirda**,
Amin Kharraz***\\
\thanks{* Corresponding Author,
\href{mailto://sadatakhavani.s@northeastern.edu}{sadatakhavani.s@northeastern.edu},
Northeastern University, Boston, MA}
\thanks{** \href{mailto://ek@ccs.neu.edu}{ek@ccs.neu.edu}, Northeastern University, Boston, MA}
\thanks{*** \href{mailto://mkharraz@fiu.edu}{mkharraz@fiu.edu}, Florida International University, Miami, FL}
}

\UseRawInputEncoding
\maketitle

\input{abstract}

\input{introduction.tex}
\input{background.tex}
\input{related_work.tex}
\input{methodology.tex}
\input{evaluation.tex}
\input{conclusion.tex}

\input{acknowledgment.tex}

\bibliographystyle{IEEEtran}
\bibliography{main}

\end{document}

%% file: abstract.tex
\begin{abstract}
Online tracking is a widespread practice on the web with questionable ethics, security, and privacy concerns. While web tracking can offer personalized and curated content to Internet users, it operates as a sophisticated surveillance mechanism to gather extensive user information.
This paper introduces PriveShield, a light-weight privacy mechanism that disrupts the information gathering cycle while offering more control to Internet users to maintain their privacy. PriveShield is implemented as a browser extension that offers an adjustable privacy feature to surf the web with multiple identities or accounts simultaneously without any changes to underlying browser code or services. When necessary, multiple factors are automatically analyzed on the client side to isolate cookies and other information that are the basis of online tracking.
PriveShield creates isolated profiles for clients based on their browsing history, interactions with websites, and the amount of time they spend on specific websites. This allows the users to easily prevent unwanted browsing information from being shared with third parties and ad exchanges without the need for manual configuration. Our evaluation results from 54 real-world scenarios show that our extension is effective in preventing retargeted ads in 91\% of those scenarios.
\end{abstract}

%% file: introduction.tex
\section{Introduction}
\IEEEPARstart{P}{rotecting} user privacy has never been 
a trivial task in the Web ecosystem. Despite various defense mechanisms in place to protect user privacy~\cite{same-op, csp, secure-token}, privacy leakage is still an important topic of discussion in the web community~\cite{emilee, csp-dead, leakage}. 
Data-driven digital marketing has had a significant role in the development of community concerns over user privacy. 
On the one hand, 
online marketing entities rely heavily on user information to provide customized content for each user. The effectiveness of their mechanisms in offering hyper-personalized content significantly depends on the quality and quantity of the collected data about users. On the other hand, it has been demonstrated in various cases~\cite{adinjection, mal_extension, adware} that such operations are deceptive or unfair commercial practices~\cite{khan_federal_nodate}.
That is, the privacy implications of some of those business acts have been significantly lower than the benefits they provide to users, or they have been purposefully misleading in order to influence the consumer's behavior or decisions about a product or service.

Internet Users have mixed feelings about data sharing practices across online marketing entities. For instance, surveys show that some users prefer targeted ads to random ads~\cite{Chanchary, dehling}, but when it comes to sharing sensitive information, such as medical information, they are concerned about it being shared with third-parties. At the same time,  privacy-sensitive users are completely hesitant about online tracking and the underlying practices~\cite{Chanchary, dehling, mcdonald}, which necessitates the collection of excessive data from web users.

Unfortunately, most current techniques are not designed very well to provide complete control over popular information sharing practices without affecting the user experience. For instance, some modern websites do not offer any services to users who do not allow cookies, or who have installed anti-tracking or anti-blocking extensions. Some other websites limit the user experience when these extensions are detected on the user's browser. Extension blocking has become a standard practice on the web, and more and more websites have started to engage in this behavior \cite{cc_detect_ad_block, ny_detect_adblock, publift_detect}. Disabling JavaScript is, unsurprisingly, not a viable option either because it can easily disrupt many of the modern features of web browsers, and web applications enabled by JavaScript. Furthermore, Private Window Mode~\cite{firefox_private} or Incognito Mode~\cite{chrome_incognito}, provided by modern browser vendors as a standard way to stay anonymous, are not designed to use the most useful features of modern browsers. Also, these modes do not persist data for further use, and when the browser is closed, all the data regarding that session in the incognito mode are wiped out which can negatively impact user experience. Lastly, users may also want to have their browsing history saved for a specific website. In this case, private modes do not meet the user's needs because closing a tab in a private window removes all information from the browser. Consequently, privacy-sensitive users are left with almost no viable user-friendly options that can offer complete segregation while maintaining the full features of web applications as well as browsers. 

\textbf{The core insight in this paper is that Internet users should be more in control of the data sharing process across third-parties, and they should be given a broader range of options to protect themselves from some of the unfair practices in this context if they choose to do so.} 
A viable approach is to augment modern browsers with light-weight techniques to offer complete segregation for browsing activities. Firefox has introduced the notion of \textit{multi-account containers} -- parallel and independent from our work -- to separate browsing activities using different browsing containers. This initiative suggests that offering more control over cookie syncing is a viable direction toward achieving more privacy-aware solutions without breaking or changing the core functionalities of web browsers. While the proposed mechanism works well in separating browsing activities, unfortunately, it requires \textit{extensive} human intervention. 
Setting multi-account containers is a non-trivial task, and is a manual process. In fact, users are left with a security service that introduces significant human overhead before being able to see the benefits of the proposed feature -- another great example of an inherently useful security service with a low \textit{economy of mechanism}~\cite{cisa2} or \textit{psychological acceptability}~\cite{cisa1}.

In this paper, we propose \System,\ a low-overhead mechanism that offers more granular control to Internet users over their privacy. We implemented the prototype of \System\ as a Chrome extension by creating and managing isolated profiles in an automated fashion.  That is, \System\ automatically analyzes website interactions, time spent on websites, and a users' browsing history, and creates isolated profiles to containerize clusters of websites of the same type. We show that this mechanism is quite useful in disrupting common cross-website tracking practices such as ad retargeting advertisements~\cite{wang2017display} where user information is shared across several tracking campaigns for more persistent web tracking, and ad exposure. In this paper, we make the following contributions: 

\begin{itemize}[noitemsep, topsep=1ex]
    \item We present the first browser extension that creates isolated profiles for different website groups based on a user's browsing history, website interaction, and session times. Because of these isolated profiles, online trackers are unable to use cookie-matching methods for targeted advertising and information sharing across ad exchanges. Cookie matching is the most widely-used method for targeted ads~\cite{bashir}.
    
    \item We crawl 77 websites from Alexa and Similarweb's top websites in each category to generate a variety of scenarios that result in the viewing of targeted ads. For the websites that do not have a specific category on Alexa or Similarweb, we used our own method to determine the website's category using Google's Natural Language Processing API. After this step, we evaluate our extension using the generated scenarios.
    
    \item Using crawled data, we evaluate our extension and demonstrate that \System\ is more than 90\% effective in preventing ad exchanges from sharing user information among themselves. Our results show that 91\% of known targeted ad scenarios failed to show a retargeted ad to our client that was using \System.
\end{itemize}

Our threat model assumes that the primary privacy risk comes from ad retargeting and cookie synchronization techniques used by advertisers and trackers to follow users across multiple websites. In particular, third-party entities exploit shared cookies and user data to profile users, leading to tracking and targeted advertising across the web. The goal of this paper is to minimize the risk of such tracking without disrupting the user’s browsing experience or significantly altering the browser's default behavior.

%% file: background.tex
\section{Background}
\label{sec:background}

Before probing into the specifics of our own research, we provide some background information on the online advertising and retargeting ads that inspired our study.

\subsection{Online Advertising and RTB}
Real-Time Bidding (RTB) is the most well-known method that is being used in online targeted ads~\cite{wang2017display}. It often involves tracking users across multiple websites to map them to products or services based on the advertisers' bids on individual impressions in real time. 

\begin{figure}[ht]
    \centering
    \includegraphics[width=\columnwidth]{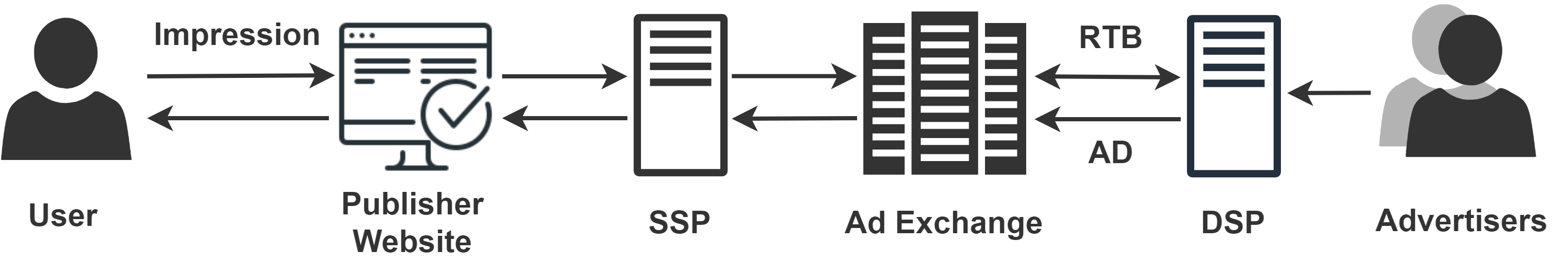}
    \caption{\textbf{High level design of the advertising ecosystem and real-time bidding.}}
    \label{fig:rtb}
\end{figure}

Figure \ref{fig:rtb} depicts a high-level design of the digital advertising ecosystem. In the ad auction process, there are typically two distinct entities, known as Supply-side Platforms (SSPs)~\cite{what_ssp} and Demand-side Platforms (DSPs)~\cite{what_dsp}. SSPs work with publishers to manage their relationships with multiple ad exchanges while focusing on revenue maximization. DSPs, on the other hand, collaborate with advertisers to determine the worth of each impression and optimize bid prices. Publishers are websites that create content and generate revenue by displaying ads to users. Advertisers are the people who wish to show ads to specific users~\cite{bashir, mayer}.

During an RTB auction, DSPs place a bid on a specific ad impression. The bid often correlates to the level of information DSPs have about a particular user. A DSP may, for example, bid higher for a known and relevant user, but it is less likely to place a high bid for an unknown user. In contrast, ad requests are routed to the SSP, which then forwards them to the ad exchange. SSPs have access to the user's cookies, while DSPs gain access to cookies only after they win the auction.

This is where cookie syncing comes into play. Using this method, the DSP could identify the user in the future when it buys ad impressions for the same user in an auction held by the same ad exchanges or SSPs. When the DSP knows who the users are, it will bid higher in order to generate more revenue.

\subsection{Cookie Syncing and Retargeted Ads}
\label{sec:cookie-sync}

Cookies are small text files that store a variety of user-related information such as email address, location, and language. 

Cookies are classified into two types: first-party cookies and third-party cookies~\cite{first_third_cookie}. First-party cookies are created by the websites visited by the user to address the stateless property of HTTP connections. 

Cookies are domain-specific, and if two different origins decide to exchange their data, the security model of the Internet browsers restricts one from reading a cookie set by another origin.
Third-party cookies, on the other hand, are created by third-party entities that are inserted into the context of a target website. For instance,
if \texttt{mynews.com} contains third-party code from \texttt{doubleclick.com}, it enables \texttt{doubleclick.com} or other dynamically loaded ads from \texttt{doubleclick.com} to create cookies that are stored on the user's browser. Third-party cookies are the foundation of digital advertising and are commonly used by advertisers and trackers.

Cookie syncing, also known as cookie matching~\cite{cookie_matching_google}, is a process that ensures advertising partners such as SSPs and DSPs can synchronize their cookies and share the incorporated user's data from their separate databases. As a result, DSPs can learn about the user's interests, demographic information, location, and so on, return the bid response with the correct bid and deliver their tailored ads to the specific target users~\cite{marjan, lukasz, acar}.

When a user navigates to \texttt{mynews.com}, multiple ad contents are often loaded on the site. This operation will create requests to the corresponding SSPs, which results in the placement of a cookie value on the user's browser containing their user ID. The SSPs also return a cookie sync pixel from their DSP partner that is called by the code executed on the web page. Cookie sync pixels are typically invisible images with a width and height of 1 pixel by 1 pixel~\cite{cookie_matching_google}. When the cookie sync pixel loads, it allows the DSP to store a cookie containing the user's User ID on the user's browser. After that, when the DSP's cookie sync pixel is called, the DSP redirects the call to an endpoint provided by the SSP. By doing this call, the DSP passes its user ID in a query string parameter to the SSP. The endpoint usually follows this pattern:  \texttt{https://ssp.com/<buyer\_id>?<user\_id>}

\begin{figure}[ht]
    \centering
    \includegraphics[width=\columnwidth]{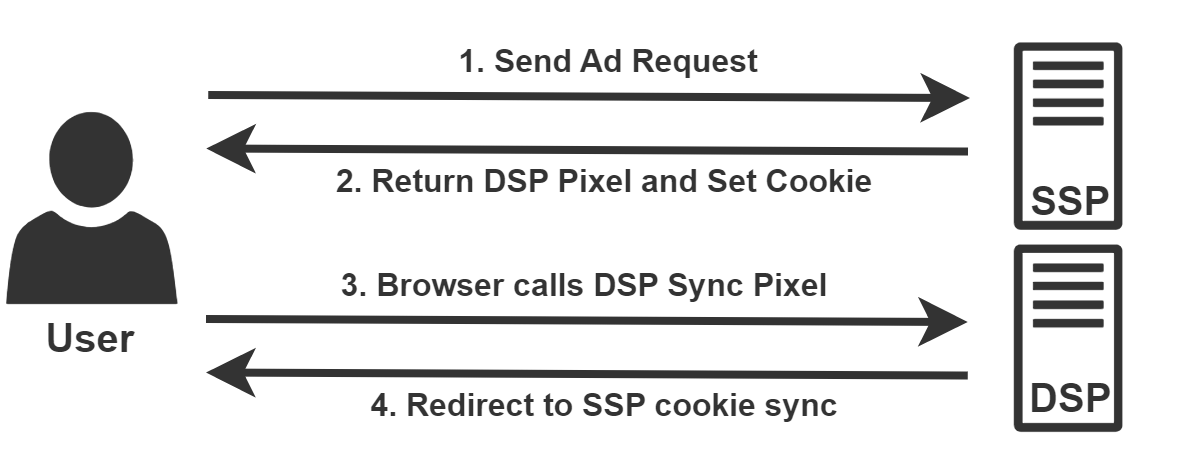}
    \caption{\textbf{Cookie syncing process between the user, SSP, and the DSP.} 1. Ad request is made to the SSP from the client's browser. 2. The SSP is able to set cookie for the user or read an existing cookie and return the DSP sync pixel. 3. The DSP cookie sync pixel gets called from the browser. 4. The DSP redirects to the SSP cookie sync endpoint to pass along their ID for the user.}
    \label{fig:cookie-matching}
\end{figure}

Here, buyer\_id is the SSP's internal ID for the DSP, and user\_id is the DSP's ID for that specific user. The SSP then reads their cookie for that user and stores their user ID alongside the DSPs user ID in a table. The DSP now has a mapping from its cookie to the SSP. In the future, if DSP participates in an auction held by the SSP, it will be able to identify matched users using SSP's cookie. Every step in the cookie syncing process between the user, SSP, and the DSP is illustrated in Figure \ref{fig:cookie-matching}.

Cookie syncing is the primary method by which ad partners can share a common understanding of identity across the web. Advertisers' targeting capabilities would be severely limited without cookie syncing in retargeted ads. In fact, targeted ads based on user data would lose their effectiveness quickly in the current ad tech ecosystem. Note that all cookies created by the SSP and DSP are third-party cookies since they will be created under the domains of those platforms rather than the domain that the user is currently visiting. On browsers that block third-party cookies, this process is not possible. 

%% file: related_work.tex
\section{Related Work}
\label{sec:related-work}

\noindent \textbf{Third Party Tracking}.
Online user tracking has been a topic of interest for researchers for significant period of time. Wang et al.~\cite{wang2017display} have done a thorough study on how digital advertising works, and discusses different RTB algorithms used by ad networks. Krishnamurthy et al.\cite{Krishnamurthy} examine the various technical ways third-party aggregators use to collect user data.
Bashir et al.~\cite{bashir} have done a study on the ad ecosystem and leverage retargeted ads as a tool for identifying information flows between ad exchanges focusing on cookie matching mechanisms. Also, Chen et al.\cite{chen} demonstrate how first-party cookies could be used in online tracking. Akhavani et al.~\cite{browserprint} examined the impact of browser features on fingerprintability and web privacy, shedding light on how browser fingerprinting affects user privacy.
Besides all these measurement studies, there have been some defenses against online user tracking across the web~\cite{pan, fran}. Also, Google is planning to phase out its support for third-party cookies in late 2024~\cite{google_third_party}.

\noindent \textbf{Content Isolation and Sandboxing}.
There have been multiple studies on how to use isolated environments so that JavaScript code related to different origins will not be able to share data across different components~\cite{op_browser, t_and_p, adjail, jsand}. By using these methods, the user will face a more secure browsing experience. These approaches focus on securing the browser and preventing malicious browsers in running attacks on the client.

\noindent \textbf{Browser Privacy Features}.
Some browsers have already recognized the need to separate cookies for specific purposes. For example, Mozilla maintains an extension known as Firefox Multi-Account Containers~\cite{multi-account_nodate}. This extension allows users to create manual profiles with cookies separated from one another. However, users must manually set their preferences, which is primarily used to keep different accounts, such as work and personal, separate from each other. Profiles are also available in Google Chrome~\cite{chrome_profiles}. Chrome profiles separate all browser settings, history, bookmarks, cookies, and even extensions in each profile. These profiles are primarily used when multiple people share a computer.

While prior research and existing browser features have made impacts in enhancing user privacy, key gaps remain in the automation and transparency of cookie separation mechanisms. Current browser solutions rely heavily on manual user intervention or are primarily focused on security rather than privacy. Moreover, approaches like sandboxing, though effective, often result in usability limitations or are detectable by ad networks, reducing their effectiveness in real-world scenarios. \System\ addresses these gaps by introducing a fully automated, privacy-focused solution for isolating browser profiles, requiring no manual user configuration. By detecting the need for profile creation and switching on its own, \System\ ensures cross-site cookie tracking is prevented without user input. Unlike existing solutions, \System\ is designed specifically to prevent retargeted ads by minimizing the background information shared with trackers, and making profile-related decisions based on this privacy goal. While some isolation techniques can be detected by ad networks or block certain website features, \System\ overcomes these challenges, filling a critical gap in defending against modern tracking methods and offering a significant advancement in user privacy protection.

%% file: methodology.tex
\section{Methodology}
\label{sec:methodology}
This paper proposes a novel technique that allows privacy-aware users to reduce the impact of cookie-syncing methods and cross-website tracking. 
To this end, we present \System, which is a browser extension that relies on automatic profiles to store and manage data from different cluster of websites in isolated profiles. The immediate outcome of using \System\ is to disrupt the cross-origin tracking cycle, and prevent ad exchanges from using cookie-matching methods to implement retargeting practices on Internet users.

\System\ is implemented using \textbf{Google Manifest V2}~\cite{google_manifest_v2}. This extension mainly targets Google Chrome, and has been fully tested on this browser. All the features and experiments discussed in this paper were analysed on Google Chrome version 98. With minor changes, we verified that the approach would also work for browsers such as Firefox as well. There has been recent discussions surrounding Google Manifest V3~\cite{google_manifest_v3} and extensions being forced to update their code base to follow the Manifest V3 pattern. \System\ has been implemented with that change in mind. Thus, migrating from Manifest V2 to V3 is straightforward, and does not affect the functionality and design goals of \System. Also, since \System's approach for profile creation does not include blocklists, migration will not affect this extension's impact on dealing with profiles. We provide all the source code and datasets that we have collected in our experiments to the community.~\footnote{\url{https://gitlab.com/sa_akhavani/priveshield}}.

\subsection{Segregation}

\System\ offers complete isolation by building a module for automatic profile generation and website assignment. That is,
by separating cookies, local storage, session storage, and cache storage for each profile, \System\ makes such critical information in one profile inaccessible to the other profiles. This approach makes it less likely for ad exchanges to use cookie-syncing methods when delivering retargeted ads. 
Consider the following scenario: A user visits Nike's website, and adds some running shoes to their shopping cart. Multiple cookies are set in the user's device as a result of this. After visiting Nike, the user decides to visit another website, such as CNN. CNN can detect if a user has previously visited Nike, and shows the user a retargeted ad from Nike, or any other company that sells running shoes by using cookie-matching. Note that this scenario could happen in a vanilla Chrome version that does not have our extension installed. In the \System\--protected environment, however, Nike and CNN belong to two different profiles. Since CNN cannot access the cookies stored by Nike on the user's device, cookie-matching methods cannot be used here. Hence, CNN would be unable to show the user a retargeted ad from Nike.

\System\  realizes the automatic profile isolation feature by introducing minimal changes to the browser settings. Our analysis shows that it has minimal impact on the user experience, and perform all the necessary tasks in the background without requiring user input. That being said, \System\ also offers a service to create a manual profile for specific websites if she decides to do so. That is, the extension also allows to create temporary profiles that delete all information after the browser is closed. The browser extension makes use of the following permissions: \texttt{history, browsingData, cookies, storage, and webNavigation}. All information is processed locally, and no data from users is sent anywhere via the Internet.
Because this extension is implemented in a modular fashion, more features may easily be added to it if so desired.

When switching between profiles, \System\ ensures that the cookies of each profile are stored separately in isolated storage. When a new profile is loaded, the cookies related to that profile are retrieved from storage and applied to the browser session. Simultaneously, the cookies from the previously active profile are updated in their respective storage to reflect any changes made during the browsing session. This process allows profile switching without any risk of cookie leakage between profiles, ensuring that the isolation is maintained across all browsing activities.

\subsection{Extension Profiles}
The core structure of our extension is built upon customized profiles. Each profile in our extension has a profile name, a profile id, related websites, cookies, local storage, and session storage. These are generated by the extension. We define 6 different profile types in \System\ that are explained briefly as follows:

\noindent \textbf{Regular Profiles}. Such profiles are generated automatically based on the user's browsing history. When a specific website is visited on a regular basis based on the user's browsing history, the extension creates a regular profile for that specific website. Each visit is defined as a distinct page view of a website. This is explained in more detail in section \ref{user-bh}.

\noindent \textbf{Session Profiles}. Such profiles are created automatically based on the amount of time a user spends on a website. If a user spends enough time on a website that requires a session profile as determined by our extension, we create a new profile for that website. This time is measured only when the website's tab in the user's browser is active and visible. This is explained in detail in Section \ref{active-session}.

\noindent \textbf{Interaction Profiles}. Such profiles are generated automatically as a result of a user's interaction with a website. If the user visits a website and interacts with that websites, for example by logging in or adding multiple products to the shopping cart, the extension recognizes that a specific interaction profile should be created for that website. Note that this is critical in order to be able to distrupt retargeting ads because 
user interactions and active session times are among the most common practices that trackers use to detect specific user behavior on tracked websites. For example, there are many services such as Plumbr~\footnote{\url{https://plumbr.io/}} and Resci~\footnote{\url{https://retentionscience.zendesk.com/hc/en-us/articles/360001874554-User-interaction-events-101}} where the aim is to detect specific kinds of user behavior and interactions such as items being added to users shopping carts, searches being conducted, item page views, etc. These actions could lead to cookies being stored on the user's browser that could then be used for tracking. 
This is explained in more detail in Section \ref{interaction}.

\noindent \textbf{Category Profiles}. Such profiles are pre-defined based on Similarweb's top website categories. When a user visits a website that does not fit into a regular, interaction, or session profile, our  extension attempts to determine the website's category. If the visited website falls into a predefined category, it is added to the matching category profile. We provide further information for these profiles in section \ref{website-category}.

\noindent \textbf{Temporary Profiles}. Such profiles do not persist cookie and storage information. When a browser is closed, all data associated with these temporary profiles will be deleted. Users can manually enable temporary mode in the extension. When temporary mode is enabled, all browser tabs open in a new isolated temporary profile. This feature, which is built on top of our isolated profiles, works similarly to the private browsing modes of browsers but does not require opening a new browser window or disabling existing extensions.

\noindent \textbf{Manual Profiles}. Such profiles are manually-created by the user, and include specific websites based on the user's needs. Users can create a new profile that includes single, or multiple websites. When navigating to a website, the extension first checks to see if there is a manual profile assigned to our target website; if there is, the website is opened in that profile.

\subsection{Profiles in Action}
Cookies, local storage, session storage, cache storage, and application cache are all unique to each of these profiles. As a result, there is no trivial way for a website in one profile to access information in another. This is the intuition behind the approach we are proposing, and which prevents cookie-matching techniques from tracking users.

The extension defines multiple event listeners, each of which is in charge of one or more of the extension's features. When a user enters a URL, an \texttt{onBeforeNavigate} event listener checks all existing profiles to see if the target website is already present in one of the profiles in the profile catalogue. If the extension discovers a profile associated with the target website, it first backs up all information belonging to the current active profile to its associated storage, and then loads all persistent information of the target profile in the browser, including cookies, session, and local storage. After all of this information has been loaded, the browser is ready to visit the target website, and navigate to the given URL.

\noindent \textbf{One Website, Multiple Profiles.} Based on the above definitions and metrics, a website might be put into multiple profiles. For example, \texttt{A.com} could be put into a regular profile \texttt{Regular-A.com} because it has been visited by the user multiple times before, and also the user might have had lots of interaction with this website so there might also be a session profile called \texttt{Session-A.com} which \texttt{A.com} also belongs to it. Whenever a user decides to visit \texttt{A.com} in their browser, \System\ looks for all profiles that contain \texttt{A.com}, and then picks the suitable profile based on the defined hierarchy. The hierarchy is as follows: 1-Manual profiles, 2-Interaction Profiles, 3-Regular Profiles, 4-Session Profiles, 5-Category Profiles.
Manual profiles have the highest priority since they are specifically created by the user. Interaction profiles are the second important ones as they are the profiles with the highest amount of cookies stored on the user side. The regular profiles have the next priority as they are being used by the user regularly, and have a high chance of storing a tracking cookie on the user side at some point in time. The last priory in this hierarchy are the session profiles. If the website belongs to none of these stated profiles, the extension looks for a category profile, and tries to open the website in a category profile.
Each regular, interaction, or session profile created by PriveShield is uniquely named and isolated for each website. For example, if both a regular and interaction profile are created for \texttt{B.com}, they would be named \texttt{Regular-B.com} and \texttt{Interaction-B.com}, which are distinct from \texttt{Regular-A.com} and \texttt{Interaction-A.com}. Since the profiles for \texttt{A.com} and \texttt{B.com}, are separate and uniquely named, the cookies and data stored in these profiles are kept isolated from one another, preventing cross-profile cookie synchronization.

\begin{figure*}[ht]
    \centering
    \includegraphics[width=0.80\textwidth]{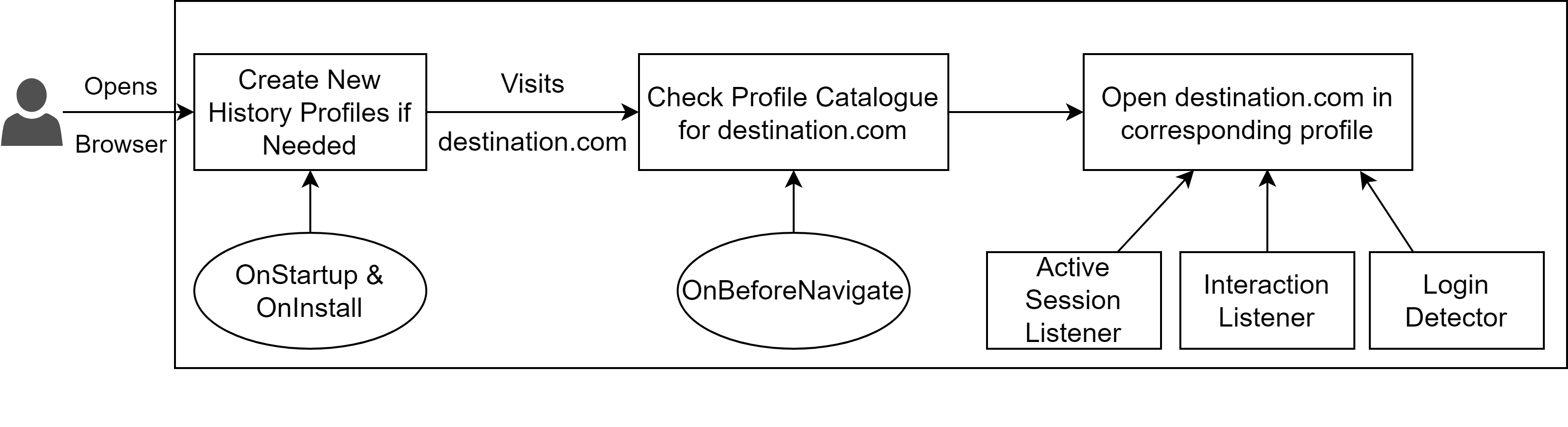}
    \caption{\textbf{How \System\ impacts the browsing process.} What happens when user opens the browser, and tries to visit \texttt{destination.com}.}
    \label{browsing-scenario}
\end{figure*}

\noindent \textbf{Default Setting.} There is a default manual profile included with the extension. This profile is in charge of storing all information related to websites that do not belong to any other profile. As a result, the first time a user visit a website, the page will be opened in the default profile. Alternatively, if the extension cannot find an associated profile for the target website in the profile catalogue, the default profile will be used. All new tabs will be opened in the default profile until the user visits a website that belongs to an existing profile, or the extension detects that the website needs to be assigned to a specific profile, and changes the profile to the corresponding accordingly. If the temporary mode is enabled, it makes no difference whether the website that the user is attempting to access belongs to a specific profile or not. All tabs are assigned a newly created temporary profile while this mode is active, and after closing each tab, all stored content related to that tab is wiped out. Temporary mode should be enabled manually and has nothing to do with automated profile detection. However, it improves the extension's usability. Figure \ref{browsing-scenario} depicts the entire browsing scenario, from start to finish, when a user attempts to visit a website.  

\subsection{User Browsing History}
\label{user-bh}
Regular profiles, as mentioned in the previous section, are generated based on the user's browsing history. To generate regular profiles, \System\ uses the user's browsing history in two different scenarios. These two scenarios are \texttt{onInstall} and \texttt{onStartup}.

\System\ collects the user's browsing history from the previous week every time the browser starts (\texttt{onStartup}), and also upon extension installation (\texttt{oninstall}) , and then extracts the number of visits and the type of visit (whether it was entered on the address bar or was clicked on a link) related to each hostname's browsing history. Our goal here is to create a dedicated profile for a hostname if we detect that the user visits the website on a regular basis.

We decided to look at the Alexa~\footnote{\url{https://www.alexa.com/topsites/countries/US}} top websites to find a suitable threshold for creating a dedicated profile for a hostname. Alexa provides daily pageviews per visitor for its top websites. We used this data to determine a weekly threshold for defining a regular website. The average daily pageviews per user for the Alexa top websites in the United States is 6.45 times per day. Thus, a regular website is one that the user visits more than 42 times per week. After installation, \System\ examines the user history, and creates a dedicated profile for a hostname if it detects that it is a regular website, which means that it has been visited more than 42 times in the previous week. The extension also includes an \texttt{onStartup} event listener, which helps in the creation of profiles based on user history. Every time the user opens the browser, it collects the user's browsing history from the previous week, and attempts to find new regular websites. If it discovers any new regular websites, it creates a new profile for them. Also, if it detects that a current regular profile is no longer needed for a website since the website is not being visited by the user regularly, \System\ deletes the regular-profile associated with that website.

This process begins after the extension has been installed. As a result, before the user visits any website, the extension creates multiple isolated profiles for the user based on their browsing history. It does not take days for the extension to study the user's behavior. Following installation, \System\ relies on the user's browsing history, and creates multiple isolated profiles for regular websites. As mentioned before, \System\ performs all of the work on the client's browser side, and no information is sent to a remote server. As a result, the user does not need to be concerned about the potential of the leakage of user personal information.

\subsection{Active session}
\label{active-session}
Active session time is another factor considered when creating automatic profiles for websites. When a user spends a specific amount of time on a single website, this could be an indication of data persistence. That is, spending more than the average amount of time on a website could indicate that the user is interacting with the website, or reading something on the website. 

As we discussed in Section~\ref{sec:related-work}, data persistence implies that a website may employ cookies to deliver targeted ads. As a result, we intend to detect and prevent cookie-syncing methods by creating isolated profiles for websites where the user has long active sessions.

We use previous research on web browsing behavior by Chao Lio et al.~\cite{dwell-time} as well as data from Similarweb's most visited websites~\footnote{\url{https://www.similarweb.com/top-websites/}} to define a threshold for active session time. Similarweb's data provides the average amount of time users spend on a website per visit for its top websites. We used this data to discover that the average visit duration among all of its top websites is 68 seconds, implying that the amount of time a user spends on a website is approximately 68 seconds.

Taking this into account, when a user has an active session on a website, \System\ calculates how much time the user has spent on that website. If the active session time exceeds our predefined threshold, the extension detects that a new dedicated profile for that website is required, and creates it. From that point forward, that website's information will be stored in its dedicated isolated profile, and future visits to the website will use the newly created profile.

\subsection{Interaction}
\label{interaction}
User involvement on websites can be used as an indicator of data persistence on visiting websites. We create a new isolated profile for a given website if we detect that the user is interacting with the website services. The interaction could occur for various reasons, including logging process, adding items to a shopping cart, looking for a product, or simply browsing different pages of a website.

\noindent \textbf{Detecting Interaction.} \System\ detects user involvement by listening to JavaScript events on the client side to determine when the user is interacting with the page. The extension detects interactions such as mouse clicks and typing on the page using the \texttt{MouseEvent}, and \texttt{KeyboardEvent} APIs. If \System\ detects more than five of these interaction events on a single web page, it establishes a new isolated profile for the host website, and triggers an event. To determine a usable threshold for number of interactions, we sampled the top 5 websites from Similarweb's top 10 most popular website categories, and averaged the number of interactions required for signing in, searching, and adding things to the user's shopping cart, -- leading to 50 different measurements. In the websites investigated, for example, a simple search scenario contained at least three interactions. Clicking on the search box, typing the target term, and pressing the enter key on the keyboard, or clicking on the search button is the most typical search scenario. This is the bare minimum of actions required in a search scenario. Other cases, such as checking in or adding items to a shopping cart, typically necessitated more interactions. However, because we were looking for a threshold, we averaged the number of interactions across all of the websites in our analysis. 

\noindent \textbf{Sensitive Operations.} Apart from counting interactions, whenever the extension detects a user logging in or signing up for a website, it creates an isolated profile for that website right away because logging in and signing up are indicators of data persistence in the browser. \System\ accomplishes this by scanning form submissions that contain \texttt{username} and \texttt{password} fields or texts related to login and sign up. This detection method is complimentary to the number of interactions method that we discussed in the previous paragraph. By using these two detection mechanisms in the browser, there is a very low probability that we will miss a user login. Unlike browsing history profiles, which are created \texttt{onStartup} and \texttt{onInstall}, active session profiles and interaction profiles are created during runtime, or in other words, at any point during the user browsing experience if \System\ detects a new profile is required. 

\noindent \textbf{Profile Updates.} Users' preferences and behavior might change over time. A specific user might meet the criteria to associate a site to a profile for a time period, but months later, may not meet that criteria anymore. For this reason, we defined a dynamic feature for and removing unnecessary profiles, and reduce the extension overhead. To do so, \System\ checks all the profiles periodically to detect necessary changes. That is, the system goes through all the stored information related to the interaction and active session profile websites, and checks if those websites have met the minimum required interactions or session time at least once during the define time slot. If a website no longer meets the profile requirements, that website's profile would be removed from \System\ to keep the extension data updated. The default time window is preset to one month, but can be changed to an arbitrary value.

\subsection{Website Category}
\label{website-category}
Some pre-defined profiles, named \textit{Category Profiles}, exist in \System\ to provide a more smooth browsing experience while also preserving the user privacy. These category profiles include the top 10 most common website categories, such as news, streaming, encyclopedias, and so on, based on Similarweb's top categories dataset. When a user wishes to access a website, but there is no profile connected with that website, these profiles are utilised. In this case, \System\ tries to detect what website category it belongs to. If the website belongs to one of the top ten most popular categories, it will be opened in the corresponding category. If the website does not belong to one of these common categories, the extension opens that website in the default profile. These profiles make it easier for the user to receive relevant suggestions because all news websites are included in the same profile if they do not have a dedicated profile.

Two methods are used by \System\ to detect a website category. It first tries to determine the website category using Similarweb's API. However, because an ordinary account on Similarweb has constraints, the API may not return a valid result. Thus, we have implemented a second method in case the first one does not return a proper response. We wrote a script that takes the web page's text content and uses Google's NLP API~\footnote{\url{https://cloud.google.com/natural-language}} to determine the website's category. The script has restrictions in that it only works on English-language websites and does not require a login to display the website's content to the user.

%% file: evaluation.tex
\section{Evaluation}
\label{sec:evaluation}
\System\ aims to add a new layer of privacy protection to the browser so that users face fewer privacy concerns while surfing the web on a daily basis. This section explains how \System\ protects users against cookie-syncing methods. \System\ is evaluated using two different scenarios. First, we define a variety of interaction scenarios that result in delivering personalized ads on a vanilla browser. We capture all of the ads that were displayed to the user after defining all of those scenarios and visiting all of the websites in a regular browser. Next, with \System\ installed on the browser, we go over all of those scenarios again. The results are then compared to determine if using \System\ can prevent delivering retargeted ads. In the second method, we analyse the stored third-party cookies in our system while using a vanilla browser and a browser with \System\ installed. Using the generated retargeted ad scenarios that we have found in the first approach, we take a deep dive into the stored third-party cookies, and compare the difference between the stored third-party cookies in each profile in \System\ to the stored third-party cookies in a vanilla browser. This second approach is performed using the concept of the tangle factor that is defined in related work, by Hu et al.~\cite{tangle-web}

By design, \System\ will prevent advertisers from showing retargeted ads to users because each tracker has different cookie instances in each profile. These cookies will contain different user IDs since trackers will assume that each profile belongs to a different user because it is separated from the other profiles. As a result, when the SSP tries to look in its match table, there may be multiple rows in the match table for a specific user due to the profiles, but they are only aware of one of them. Thus, cookie syncing between profiles will be ineffective. In the following subsections, we evaluate the performance of \System\ in real-world scenarios while using the extension on existing websites.

All the evaluations and crawling in this research are done using Google Chrome version 98. Also, Selenium version 4.0.0 and Puppeteer version 10.2.0 have been used for automated testing and performance evaluations. The test machine configuration includes 8GB of RAM, Intel Core i5-10210U CPU with four cores, and Ubuntu 18.04 as the operating system.

Both the vanilla version and the \System\--enabled simulations were run in parallel at the same time and visits to the same website had identical interactions and active session time to prevent a difference in the data persistence of target websites in these two scenarios.

\subsection{Usability Testing}
\label{sec:eval-1}
To evaluate the effectiveness of \System\ on users' daily activities, we conducted a manual evaluation on top websites that displayed ads in Similarweb's top categories. The evaluation is designed to demonstrate that users will not receive personalized retargeted ads on different website categories based on their searching history, active sessions, website interactions, or shopping wish lists if they use our extension. In this subsection, we run a simple scenario through two browsers. One browser has \System\ enabled, while the other does not have our extension installed.

\noindent \textbf{Data Collection.}
We define two different website classes. One class includes \emph{e-commerce} websites. These are the websites that typically sell a product and wish to have their ads displayed to users on other websites. These e-commerce websites are divided into 14 categories, including running\_shoes, jewelry, cars, travel, computer\_equipment, and so on. We collected 40 different English-language e-commerce websites in total. The complete list of e-commerce website categories is depicted in Table \ref{table:1}. The other website class contains \emph{publisher} websites. These websites do not usually sell products, and instead, generate revenue by displaying ads related to the e-commerce website on their website. These websites may also fall into various categories such as news, sports, finance, and so on. We collected 37 publisher websites in total, all of which were in English, did not require login to view content, and were not in the gambling or adult categories.

\noindent \textbf{Evaluation.} One question we wanted to answer first was to identify e-commerce and publisher websites that show retargeted ads to their users using a vanilla browser. To this end, for each e-commerce category, we visited all the websites in our dataset that belonged to that category. Then for each of those websites, we visited 4 different pages on that website to increase the chance of seeing retargeted ads related to that product category. After visiting these pages on the e-commerce websites, we went through all of the publisher websites in our dataset one by one. For each of those publisher websites, we visited it twice, and took a screenshot each time. This whole process was automated using Puppeteer and custom scripts. Then, we manually went through the screenshots to see if we had shown a retargeted ad related to the original e-commerce category or not. Note that we visit each website in the publisher category twice to ensure that the ad we are seeing is not related to the e-commerce website at random. We save the scenario that results in a retargeted ad if we see it in both screenshots. Following these runs, we discovered 10 e-commerce website categories that resulted in retargeted ads in our tests out of the 14 different e-commerce website categories. These categories are shown in table \ref{table:1}. In addition, 24 publisher websites (out of 37 in our dataset) displayed a retargeted ad in at least one scenario. In total, we identified 54 distinct scenarios that resulted in retargeted ads. 

\begin{table}[t!]
\centering
\caption{\textbf{Evaluation results for each website category.} The table depicts which website categories in our study displayed retargeted ads to our client before using our extension. The third column provides the number of publishers that displayed a retargeted ad related to the corresponding category.}\label{table:1}
\begin{tabular}{l|c|c}
\hline
\multicolumn{3}{c}{Retargeted Ad Scenarios} \\
\hline\hline
Category Name & Retargeted ad? & Number of Publishers\\
\hline
running\_shoes  & yes & 9\\
 clothing   &   yes & 2\\
 watch   &   yes & 3\\
 jewelry   &   yes & 6\\
 cars   &   yes & 4\\
 computer\_equipment   &   yes & 8\\
 banking   &   yes & 7\\
 insurance   &   no & 0\\
 streaming   &   yes & 4\\
 marketplace   &   no & 0\\
 home\_decor   &   yes & 8\\
 health   &   no & 0\\
 travel   &   yes & 3\\
 rental\_cars   &   no & 0\\
 \hline
\end{tabular}
\end{table}

\noindent \textbf{Real-World Scenario Usage.}
We used these real-world scenarios that led to retargeted ads to measure the effectiveness of \System\ when it is installed on a browser. We manually visited the 4 custom URLs on the e-commerce website related to each of the 54 scenarios. These URLs were the same pages that were visited in the previous step to generate the retargeted ads. These page visits could automatically trigger the event listener in \System, and the extension would create a profile for that e-commerce website. After that, we visited the publisher website related to that scenario two times and took a screenshot of it. Finally, we examined the screenshot to determine whether or not we still had a retargeted ad. Our results showed that retargeted ads were not observed in 49 of the total 54 scenarios while using \System. This means that our extension was 90.74\%  effective at preventing retargeted ads. The five scenarios in which we saw retargeted ads could be due to a variety of factors. Although cookie-syncing methods were the main focus of this work, websites might use other methods for displaying ads. Thus, these five scenarios might be caused by other advertising methods. Furthermore, because we only visited the publisher website twice, these ads may also have been shown to the user randomly.

\begin{figure}[ht]
    \centering
    \includegraphics[width=\columnwidth]{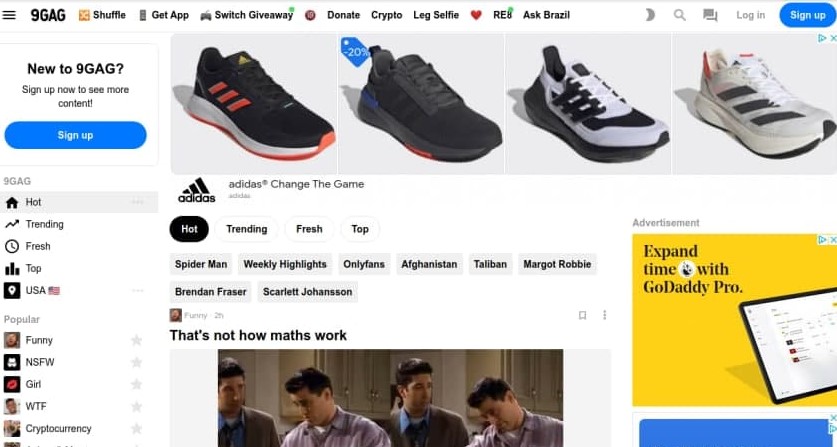}
    \caption{\textbf{Example of a retargeted ad displayed in a publisher website.} A \texttt{running\_shoe} ad is being shown on 9gag's website. This screenshot is taken from 9gag after visiting a scenario that included multiple web pages in the ``running\_shoe'' category.}
    \label{fig:retarget}
\end{figure}

\noindent \textbf{Example Scenario.} 
We use an example scenario to demonstrate how this evaluation works on one well-known e-commerce website, and one from our publisher list. We select ``adidas.com'' from our e-commerce website list, which is in the running\_shoes category. In addition, we select ``9gag.com'' from our publisher websites. In a vanilla browser, when a user visits four different URLs on adidas.com's website before proceeding to 9gag.com's website, an ad related to the running\_shoes category is displayed to the user (shown in Fig. \ref{fig:retarget}). This is one of our 54 retargeted ad scenarios. If we run the same scenario in a browser that has \System\ installed, we are no longer shown an ad related to the running\_shoes category, but rather a random ad on 9gag's website. This is one of the scenarios in which the extension successfully prevents retargeted ads. Our results confirm that \System\ is effective in reducing the likelihood of seeing retargeted ads through automatic profile generation. The real-world scenarios that were used for this purpose typically show retargeted ads to the user when they use a vanilla version for browsing. 

\subsection{Cookie Comparison}
\label{cookie-comparison}
One way to assess the effectiveness of \System\ on reducing users' exposure to retargeted ads is to measure the activity of third-parties in setting and exchanging cookies during normal web browsing. That is, 
we compare the number of third-party cookies stored in a vanilla browser during a web browsing session, and compare the value to the one when we run with a \System\--enabled browser. We follow the same steps mentioned in the previous section to interact with websites that perform cookie-syncing and collect the artifacts for evaluation to see how \System\ affects third-party cookies.

\noindent \textbf{Tangle Factor.} We incorporated a metric called\textit{tangle factor} defined by Hu et al.~\cite{tangle-web}, which is a measure that determine how first-party websites may be interconnected with other websites based on the common third-party code shared among those websites. A higher tangle factor means that a larger number of third-party cookies are interacting with each other when a user visits a page. Thus, we incorporate this measure in \System\ to evaluate the effectiveness of the proposed approach in decreasing the interaction among third-parties. We run the experiment by monitoring the tangle factor while generating automated isolated profiles, and measuring if the tangle factor decreases compared to the vanilla version.   We also used the graph representation used in~\cite{tangle-web} to show third-party interaction by incorporating tracker data as well as cookie syncing across the interconnections of trackers.

We picked two random different real-world retargeted ad scenarios generated in subsection \ref{sec:eval-1} , each from a unique category for our measurement demonstration. These two scenarios belonged to \texttt{running\_shoes} and \texttt{home\_decor}. For each of these scenarios, once in a vanilla browser and once in a \System\--enabled browser, we visited all the e-commerce websites in the corresponding category, and then visited the publisher website that was supposed to display a retargeted ad. After this, we exported all the third-party cookie information generated by \texttt{Thunderbeam-Lightbeam} in order to measure the effectiveness of \System.

\begin{figure}[ht]
    \centering
    \includegraphics[width=\columnwidth]{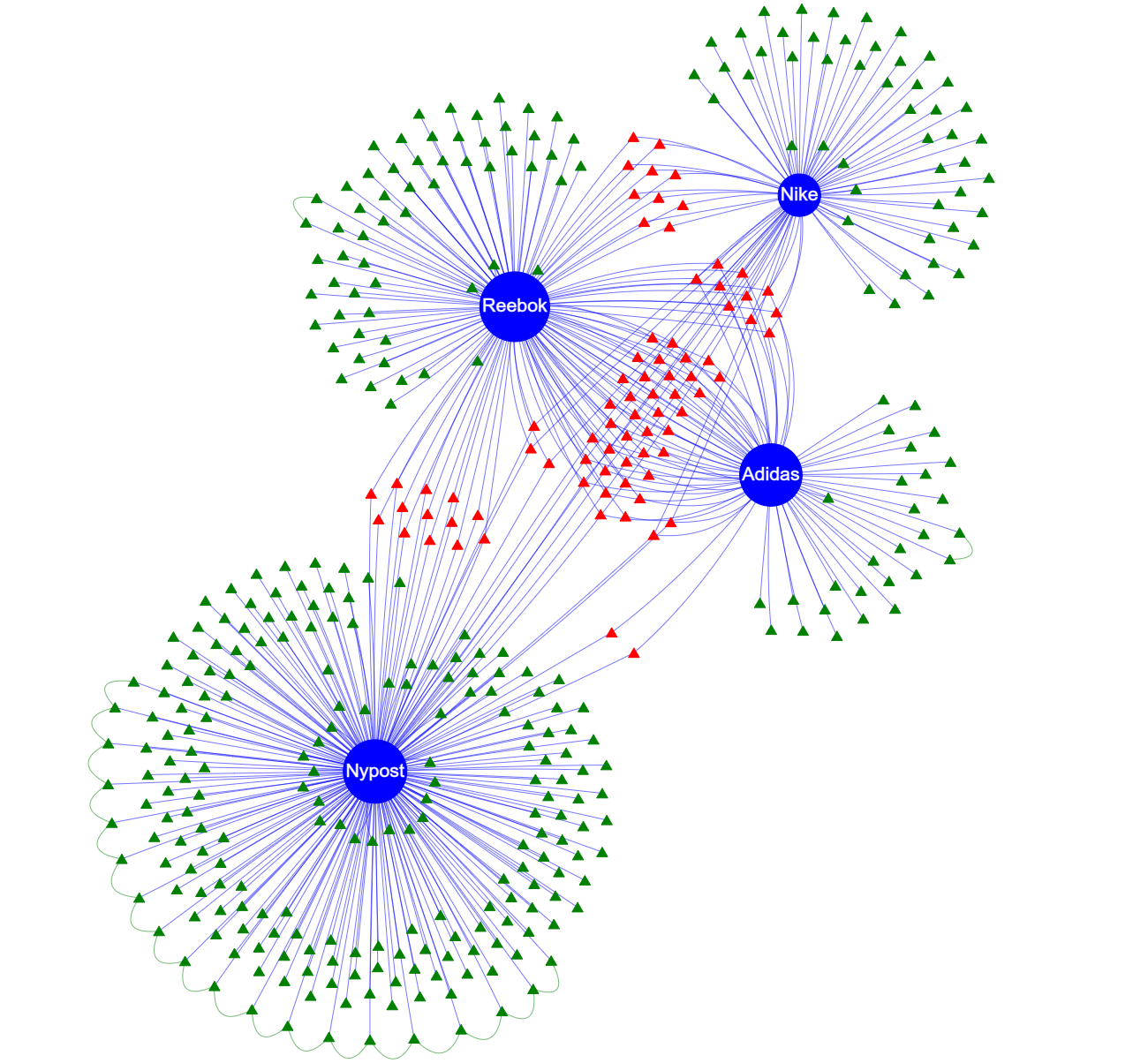}
    \caption{\textbf{Graph of first and third-party websites after visiting a scenario that leads to a running shoe retargeted ad in a vanilla browser.} Blue circle nodes are first-party websites, and the triangle nodes are the third-party trackers. Red triangles are the ones that are shared between at least two first-party websites, and are able to share information. In this scenario, the retargeted ad is shown on Nypost's website after visiting and interacting with Reebok, Adidas, and Nike's websites.}
    \label{fig:graph-running}
\end{figure}

\noindent \textbf{Tangle Factor in the Wild.} Figure~\ref{fig:graph-running} displays the exported third-party tracker data in an example \texttt{running\_shoes} scenario when we have used the vanilla browser. In this scenario, we have visited and interacted with three different e-commerce running shoe websites that are nike.com, reebok.com, and adidas.com. Then, we visited the nypost.com website which is a known publisher, and has displayed a retargeted ad belonging to the running shoes category to us. There are four clusters in the generated graph, and each cluster belongs to one of these websites and its third-party trackers. In order to prevent the communication of these trackers, in principle, we would need at least 4 isolated profiles so that the first-party websites' common trackers cannot share data among each other. This means that the tangle factor of this whole network is four. Now, we measure how \System\ reacts to the same scenario and analyse if it is able to generate some automatic isolated profiles to put these first-party websites and their trackers in them. To evaluate \System, we analysed the exported data for the exact same scenario when \System\ is installed on the browser. We observed that \System\ generated four isolated profiles for each of those websites, and put all the stored information separated from the other websites.

As a result, nypost.com no longer displayed a retargeted ad while using \System.
We also did not observe any cookie-syncing during the process. The results demonstrated that each generated profile in \System\ was no longer having multiple clusters connected to each other, and only had one cluster. Consequently, the tangle factor for all the profiles in our example scenario was one, and \System\ prevented the display of retargeted ads across those websites. 

\begin{figure}[ht]
    \centering
    \includegraphics[width=0.7\columnwidth]{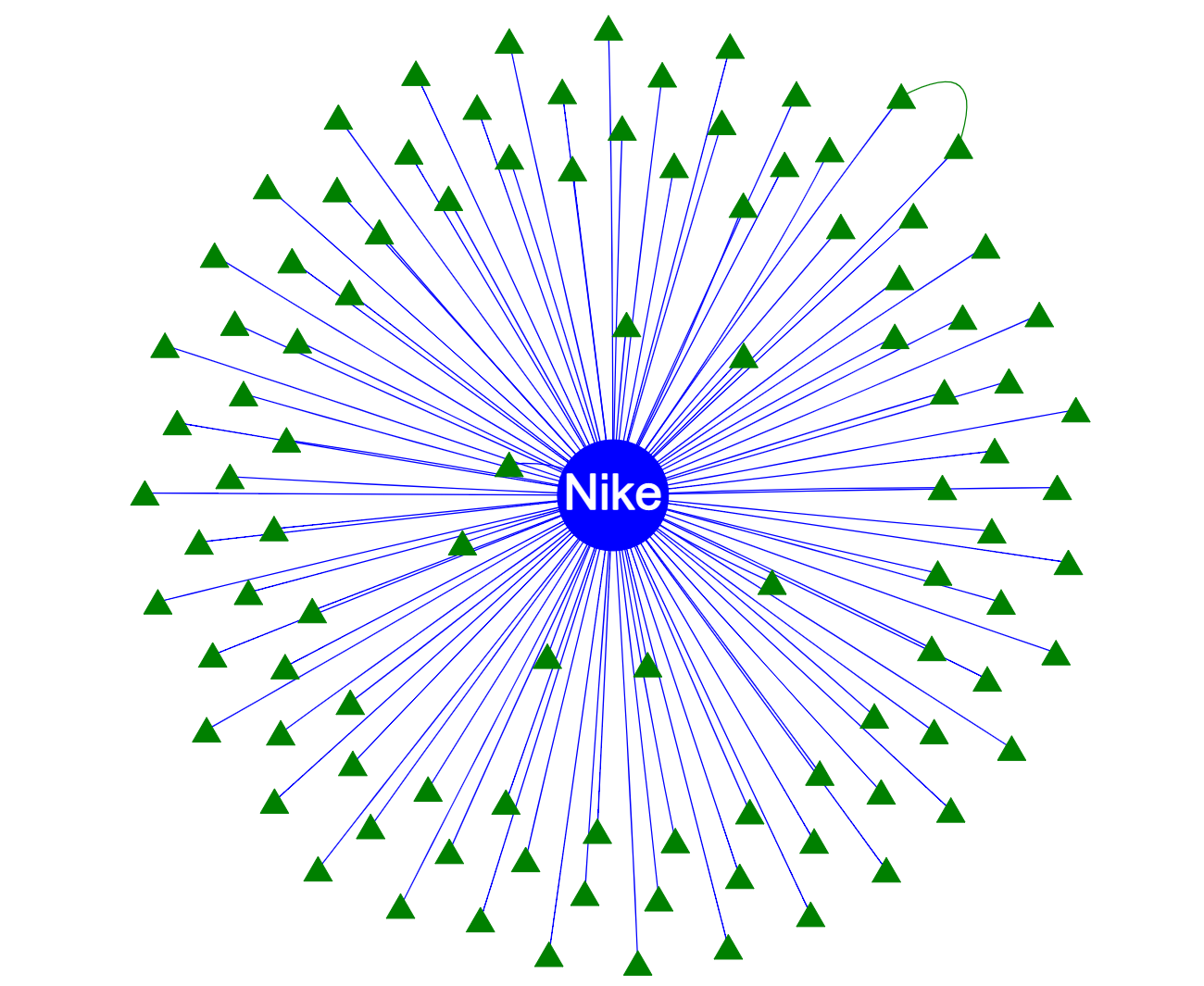}
    \caption{\textbf{Graph of first and third-party websites for the nike.com profile after visiting a scenario that leads to a running shoe retargeted ad while using \System.} As  can be seen, only nike.com and its direct trackers are stored in this profile. That is, there is no evidence of other websites and their trackers such as Reebok and Adidas.}
    \label{fig:graph-nike}
\end{figure}

\begin{figure}[ht]
    \centering
    \includegraphics[width=0.9\columnwidth]{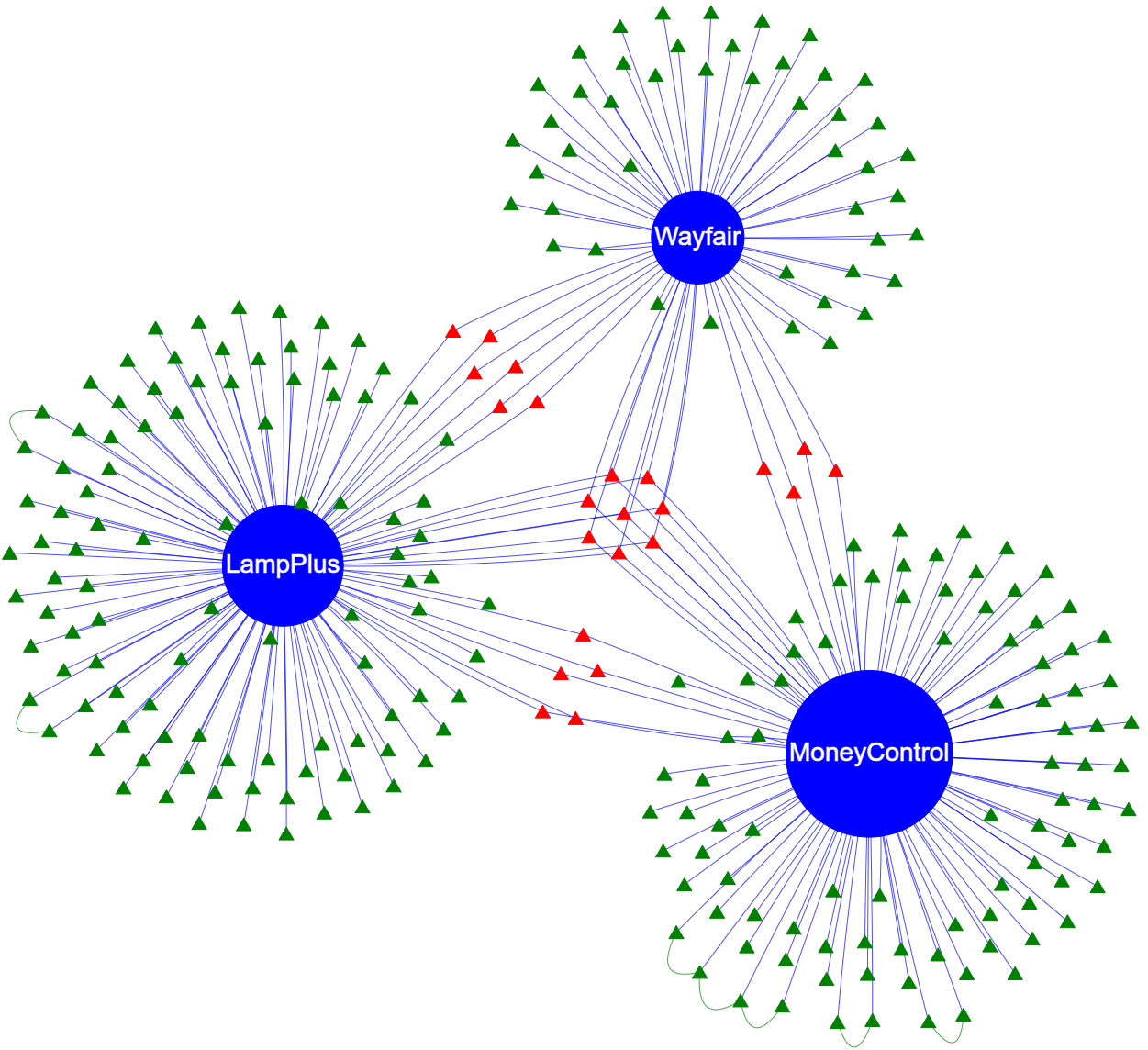}
    \caption{\textbf{Graph of first and third-party websites after visiting a scenario that leads to a \texttt{home\_decor} retargeted ad in 9gag.com's website while using a vanilla browser.} moneycontrol.com is a known publisher which is displaying the retargeted ad, and wayfair.com and lampplus.com are well-known e-commerce websites in the \texttt{home\_decor} category}
    \label{fig:graph-home}
\end{figure}

Figure~\ref{fig:graph-nike} displays the graph when the Nike profile was generated by \System\ is activated. We observe that there are no longer third-party cookies belonging to reebok.com, adidas.com, and nypost.com in the graph because they are being stored in other isolated profiles.
We also performed the same experiment on all the other scenarios that belonged to different website categories. We observed that \System\--generated isolated profiles with the tangle-factor of 1 in all of the cases which shows that cookie-syncing for ad-retargets was not allowed. Figure\ref{fig:graph-home} demonstrates the first and third-party websites in another scenario that belongs to the home\_decor category.

We demonstrated that in all these scenarios, \System\ generates some automated profiles for the websites that have been visited. When we look at the output of the exported third-party cookies data, these isolated profiles do not have shared third-party trackers inside of them. But in a vanilla browser, we observe that retargeted ads are being displayed because cookie-syncing is happening, and the tangle factor after visiting all these scenarios is high.

In summary, the first evaluation method applies the extension to real-world scenarios, and tests its effectiveness with scenarios that typically show retargeted ads to the user in a vanilla version. This method demonstrates that the isolation profile approach used in our extension is effective. The second evaluation method shows the effectiveness of \System's automatic profile generation and its ability to reduce the likelihood of seeing retargeted ads. The second approach analyses what happens to the third-party cookies in profiles while a retargeted ad is being shown, and how \System\ prevents the display of these retargeted ads.

\subsection{Performance Cost}
\System\ makes use of all of the browser's built-in functions to achieve its privacy goals. For instance, it employs a specific set of event listeners, which may cause delays in the content displayed to the user. The context switch between isolated profiles is also the most concerning aspect in terms of performance cost. In a normal scenario when \System\ is enabled, the browser behaves almost as if no extension is installed. However, when \System\ detects that it needs to create a profile, or switch to an existing profile, it attempts to store and reload cookies. Thus, we ran an experiment to see how these profile switches and event listeners impact the performance.

To accomplish this, we used the same method as in the cookie comparison subsection \ref{cookie-comparison}. In particular, we used a vanilla browser and a browser with \System\ and visited 20 sample websites 20 times, measuring the values of \texttt{requestStart}, \texttt{responseStart}\footnote{\url{https://www.w3.org/TR/resource-timing-2/}}, and the  \texttt{domComplete}\footnote{\url{https://w3c.github.io/navigation-timing/}} for all visits. We defined the \texttt{backendPerformance} as the difference between requestStart and responseStart, and also, the \texttt{frontendPerformance} as the difference between the \texttt{responseStart} and the \texttt{domComplete}. We measured the average backendPerformance and frontendPerformance of a vanilla browser and a browser with \System\ by collecting these values for all 20 websites. To make the comparison more accurate, we ran the experiment 20 times on each scenario to eliminate the effect of network latency, server response time, and other random measures. Selenium was used to automate this testing process, and measure all these navigation timing APIs.

Our findings show that \System\ has very little effect on backendPerformance. Our performance test results showed that on a normal browser, our backendPerformance average value was 180ms, while on a browser with \System, this value was 202ms. There is a time difference of 22 milliseconds. The extension's impact on frontendPerformance is also so minor that the user may not notice it. On a vanilla browser, the average value for frontendPerformance was 2035ms, while on a browser with the extension installed, the value was 2404ms. Thus, the extension has a 12\% impact on the browser's backendPerformance, and an 18\% impact on the browser's frontendPerformance.

%% file: conclusion.tex
\section{Discussion and Conclusion}
\label{sec:conclusion}

In this project, we implemented \System, a low-overhead privacy tool, to disrupt 
the intrusive tracking practices in the wild. That is, \System\
alters the way cookies are stored on the client-side, making cookie-syncing techniques significantly less effective in displaying retargeted ads to users. 
In the following, we briefly explain some of the fundamental design principles we considered in proposing \System\ , and some of the challenges we had to tackle to achieve those goals. 

\noindent \textbf{Usability.}
\System\ reuses all of the browser's built-in functions to achieve the privacy goals. This was one of the intentional choices  behind our approach. In fact,  the primary goal of this project was to show that it is possible to add new low-overhead features for users to improve their privacy without impacting their usability, changing the underlying code of the browser, or introducing intrusive functionalities to the browsing experience.  

As mentioned earlier in Section~\ref{sec:evaluation}, the empirical tests suggest that the performance impact is negligible when \System\ is integrated with the browser as the average run-time performance overhead was 202 ms compared to 180ms in the normal vanilla version.

\noindent \textbf{Work Factor.}
The economy of mechanism~\cite{cisa2}, as a security design principle,was a critical part of the design process in this project. Disrupting cooking syncing and cookie matching mechanisms, which are the core functionalities of the retargeting process, is a complex and non-trivial problem. We investigated different ways to look at this issue (e.g., modifying the Chromium source code or automatic browser feature reduction). However, all of those updates require fundamental changes in the code and underlying functionality of the browser, and it is less likely to be used by end-users. The goal of this work was to add privacy features to disrupt the ad-retargeting process by incorporating minimum layers into the browser. That is, \System\ serves a new layer of defense by introducing a form of defense asymmetry for defenders. In particular, generating new profiles does not introduce significant cost on the user-side, but at the same time, it significantly reduces the effectiveness of modern privacy leaking practices at the web scale due to the higher degree of isolation at the profile level (i.e., isolation for cookies, local and session storage, and cache). 

\noindent \textbf{Complete Isolation.} We understand that in order to achieve a completely private solution in our scenario, we need one profile for each website. However, this has a significant impact on user experience because each website only has cookies related to itself. As a result, no content can be shared between websites, third-party cookies will no longer function, a profile switch is required for each website visit, and users will no longer receive appropriate recommendations based on their previous interactions with websites other than the one they are currently visiting. Unfortunately, although straight-forward, this has a negative impact on the user experience, and is not a viable solution. \System\ attempts to find a balance between usability and privacy by employing various approaches to create dedicated profiles for specific websites as well as putting multiple websites in the same profile to provide a better user experience.

\noindent \textbf{Impact on Digital Platform Revenue.}
\System\ is designed to enhance user privacy without disrupting the revenue models of digital platforms that rely on advertisements. It is important to clarify that \System\ is not an ad blocker; it does not prevent ads from being shown to users. Instead, \System\ limits cross-site data sharing between unrelated third parties, reducing unnecessary tracking while allowing advertisements to continue functioning. As demonstrated in our evaluation, without \System\, user data is frequently shared across various entities. With \System\, data sharing remains possible within the same website category, allowing platforms to serve contextually relevant ads based on user behavior within specific contexts. This approach preserves the functionality of ad-supported platforms while enhancing privacy by preventing unrelated third parties from tracking users across multiple domains.

\noindent \textbf{Empirical Evaluation.}
Last but not least, the evaluation of results was not straightforward. Although cookie matching is the most commonly used method for online advertising~\cite{bashir}, it is not the only one. Other methods such as browser fingerprinting, or indirect matching are also used by ad exchanges to display targeted ads. Furthermore, we also needed a way to distinguish whether an ad was retargeted or random. To achieve this, we extracted all ads shown on our browser, and manually verified whether or not they were retargeted. By focusing on cookie matching methods, our extension prevents retargeted ads from being displayed to the user. Because cookies are isolated between different profiles in our extension, cookie-matching methods for ad exchanges are not possible. Using our extension, we expect to see no retargeted ads that use cookie matching methods. Cookie matching, on the other hand, is not the only method used by trackers to display targeted ads. Indirect matching methods, such as browser fingerprinting, that do not rely on cookies to identify a user, remain viable. The focus of our work was to increase the difficulty bar for cookie matching techniques, which are the most commonly used technique by trackers~\cite{bashir}.

\noindent \textbf{Future Work.}
As part of future development, we plan to explore incorporating fingerprinting avoidance techniques into \System. Browser fingerprinting, which involves tracking users based on unique characteristics of their device and browser settings, presents a growing privacy concern. Although initial experiments with fingerprinting avoidance—such as modifying request headers or randomizing certain browser attributes—showed promise in confusing trackers, these techniques require more invasive access to user data and permissions than our current lightweight design allows. Furthermore, incorporating fingerprinting avoidance would complicate the evaluation process, as it would be challenging to distinguish whether successful ad retargeting prevention was due to cookie isolation or fingerprinting interference. Future work will involve addressing these technical challenges while maintaining \System's focus on minimal browser intrusion and performance impact.

%% file: acknowledgment.tex
\section*{Acknowledgment}

This work was supported by the National Science Foundation under grants 2329540 and 2219921.